\PassOptionsToPackage{square, numbers, sort}{natbib}
\documentclass[preprint,apl]{revtex4-2}
\usepackage[utf8]{inputenc}
\usepackage[T1]{fontenc}
\usepackage[english]{babel}

\usepackage[fleqn]{amsmath}
\usepackage{mathtools}
\usepackage{amssymb}
\usepackage{xfrac}
\usepackage{physics}

\usepackage{placeins}
\usepackage{graphicx}
\usepackage{caption}
\usepackage[percent]{overpic}
\usepackage[breaklinks]{hyperref}
\usepackage{booktabs}
\usepackage{breakurl} 
\usepackage{tabularx}
\usepackage[graphicx]{realboxes}
\usepackage{array}

\usepackage{xcolor}

\newcommand{\ii}{\mathrm{i}}
\newcommand{\ee}{\mathrm{e}}

\usepackage{siunitx}

\begin{document}

\title{Signatures of real-space geometry, topology, and metric tensor in quantum transport in periodically corrugated spaces}
\author{Benjamin Schwager}
\author{Theresa Appel}
\author{Jamal Berakdar} 
\affiliation{Institut für Physik, Martin-Luther-Universität Halle-Wittenberg, 06099 Halle, Germany}
\date{\today}
\begin{abstract}
The motion of a quantum particle constrained  to  a two-dimensional non-compact Riemannian manifold  with non-trivial metric can be 
described  by  a flat-space Schrödinger-type equation at the  cost of introducing local mass and  metric and geometry-induced effective potential with  no classical counterpart. For a metric tensor  periodically modulated along one dimension, the formation of bands is demonstrated and transport-related quantities are derived.
 Using S-matrix approach, the quantum conductance along the manifold is calculated and contrasted with conventional quantum transport methods in flat spaces. 
{  The topology, e.g. whether the manifold is simply connected, compact or non-compact shows up in global, non-local  properties such as the Aharonov-Bohm 
 phase}.  The results vividly demonstrate  emergent phenomena due to the interplay of reduced-dimensionality, particles quantum nature, geometry, and topology.
\end{abstract}
\maketitle
%
%
{\it Introduction:-}The manifestations of  the underlying  geometry and topology of a physical system in its  behavior have been discussed recurrently and in varying contexts \citep{Pancharatnam1956, Karplus1954, Thouless1982, Berry1984, Aharonov1987, Samuel1988, Volovik2003, Bernevig2006, Frankel2012, Cohen2019}. For a periodic system, for instance,  the single particle spectrum is conventionally described  in the wave-vector space (or $\mathbf{k}$-space) and is characterized by bands. The geometry of the $\nu$-th energy band, $\nu \in \mathbb{N}$, viewed  in the parameter $\mathbf{k}$-space \citep{Zak1989, Xiao2010} is captured by the so-called quantum geometric tensor $T_\mathbf{k}^{\nu}$ \cite{Provost1980, Matsuura2010, Kolodrubetz2013, Ma2013, Yang2015, Bernevig2013}. Using the projector onto the $\nu$-th energy manifold, $\hat{P}_\mathbf{k}^{\nu} = |u_\mathbf{k}^{\nu} \rangle \langle u_\mathbf{k}^{\nu}|$ where $\vert u_{\mathbf{k}}^{\nu} \rangle$ is the Bloch state representing the space-periodic amplitude $u_{\mathbf{k}}^{\nu}$ indexed by the parameter pair $(\nu; \mathbf{k})$, the quantum geometric tensor reads
$    (T_{\textbf{k}}^{\nu})_{ij} = \langle \partial_{k^{i}}\,u_{\mathbf{k}}^{\nu} \left\vert \hat{I} - \hat{P}_\mathbf{k}^{\nu} \right\vert \partial_{k^{j}}\, u_{\mathbf{k}}^{\nu} \rangle = (g_{\mathbf{k}}^{\nu})_{ij} + \frac{\ii}{2}\,(\Omega_{\mathbf{k}}^{\nu})_{ij}.
$
The (symmetric) real part $\mathrm{Re}(T_{\textbf{k}}^{\nu})_{ij} = (g_{\mathbf{k}}^{\nu})_{ij}$ has the meaning of a metric  associated with the $\nu$-th energy manifold $M_\nu$.  The (anti-symmetric) imaginary part $\mathrm{Im}(T_{\textbf{k}}^{\nu})_{ij} = \frac{1}{2}\,(\Omega_{\mathbf{k}}^{\nu})_{ij}$ is proportional to Berry's curvature. Technically, this metric allows for  a notion of distance
$ (\mathrm{d}s^{\nu})^{2}$: For two Bloch states in the manifold $M_\mathbf{k}^\nu$, one may write
$
    (\mathrm{d}s^{\nu})^{2} = 1 - \left\vert \langle u_{\mathbf{k}}^{\nu} \vert u_{\mathbf{k} + \mathrm{d}\mathbf{k}}^{\nu} \rangle \right\vert^{2} = (g_{\mathbf{k}}^{\nu})_{ij}\ \mathrm{d}k^{i}\,\mathrm{d}k^{j}.
$
Coherent transport in linear response to a driving external spatially homogeneous  weak vector potential can be formulated in terms of Berry's curvature, which can be viewed as a tensor gauge-field derived from  Berry's vector potential (connection) $\boldsymbol{\mathcal{A}}_{\mathbf{k}}^{\nu} = \ii\,\langle u_{\mathbf{k}}^{\nu} \vert \boldsymbol{\nabla}_{\mathbf{k}} \vert u_{\mathbf{k}}^{\nu} \rangle$, in which case Berry's curvature reads $\boldsymbol{\Omega}_{\mathbf{k}}^{\nu} = \boldsymbol{\nabla}_{\mathbf{k}}\times \boldsymbol{\mathcal{A}}_{\mathbf{k}}^{\nu}$.
\par
The concept of Berry's curvature has proven highly illuminating and resulted in a unified view on the role of geometry in effective single particle physics. In this example the system exists in real (flat) Euclidean space and the quantum geometric tensor, and therefore associated phenomena, are determined by the symmetry class of the crystal only. The concept of  the $\mathbf{k}$-space
Riemannian manifold $(M_{\mathbf{k}}^\nu, g_{\mathbf{k}}^\nu)$ relies on the existence of well-defined (infinitely long-lived) modes at certain $\mathbf{k} \in \mathbb{R}^{3}$, a condition which in reality is compromised by the presence of electronic correlations, disorder, or other $\mathbf{k}$-states mixing effects. In addition, near thermal or quantum critical phase transitions the quantum metric generally diverges and one has to deal with issues of diverging susceptibilities and singular manifolds. Variations of Berry's curvature and the quantum metric tensor due to small deviations from the scenario sketched above, e.\,g. due to disorder, can be dealt with via perturbation theory \citep{Kolodrubetz2013, Piechon2016, Chen2022, Liu2024, Gao2023, Parker2019}, basically along the lines of  treating perturbations beyond Kubo's linear response functions.
\par
In a more general context, the evolution of a quantum system can be geometrically characterized via  the quantum distance $L(t,t_0) = \int_{t_0}^{t} 1\ \mathrm{d}s(t')$, where $\mathrm{d}s(t') = 2\,\sqrt{1-\vert \langle \Psi(t') \vert \Psi(t'+\mathrm{d}t')\rangle\vert^{2}}$ is the infinitesimal
distance as measured by the Fubini-Study metric \citep{Boya1989}. Thus, the evolution is viewed as a curve in the Hilbert space \citep{Pati1995, Aharonov1987, Cohen2019}. The state $\vert\Psi\rangle$ can be correlated or field driven. The Bargmann angle \citep{Boya1989} $\beta(t)$, with $\cos(\beta(t))=\vert \langle \Psi(t)\vert \Psi(t_0)\rangle\vert$, is a measure for the deviation of the initial state at $t_{0}$ from the target state at $t$. Thus, it is related to the fidelity \citep{Gu2008, Yang2008, Albuquerque2010, Gu2010} and can be used to find conditions for dynamical localization and even freezing, meaning a quasi-stationarity of states even in the presence of external driving \citep{MatosAbiague2006, MatosAbiague2005}. These ideas and methods can be applied also in the case discussed below, once we determine the basis of the  Hilbert space.\\ 
\par
A further  setting where geometry and topology  imprint effects on quantum motion is when the real-space manifold in which the states live
is equipped with a  nontrivial (spatially varying) metric. This situation is typically encountered when formulating quantum mechanics under spatial constraints.  The particle moves then according  to the  Schrödinger equation within a Riemannian manifold embedded in a higher-dimensional  space (hyperspace). In addition to electronic systems in reduced/strained geometries such cases may also occur in magnetic, mechanical, and optical systems. What seems as a straightforward extension of classical Lagrangian formalism  for constrained   motion to the quantum domain is in fact highly non-trivial, and  different treatments  are under debate; one of which is the confining potential approach (CPA) \citep{Koppe1971, Jensen1971, Costa1981, Costa1982, Costa1986, Maraner1995, Schuster2003, Maraner2008, Brandt2017} that shall be considered here. In this method,   the particle resides on the manifold in  a narrow region around   hyperspace-manifold interface, as dictated by Heisenberg-uncertainty. Its  quantum dynamics is shown to be sensitive to the geometry of the underlying space via so-called geometry-induced effects arising as corrections to quantum operators \citep{Wang2017}, most prominently a scalar potential-like term. 
\par
In the context of the CPA,  studies on periodically corrugated spaces can be classified according to their topology as compact \citep{Encinosa1998, Cheng2018, Cheng2019, Torres2019, Pereira2024} or non-compact \citep{Ono2009, Ono2010, Santos2016, Wang2016, Cao2019, Serafim2019, Zhao2021, Serafim2021}. In the compact case periodic boundary conditions are enforced as the canonical choice for the solution of the effective Schrödinger equation, implying, due to uniqueness of wavefunction,  a discrete spectrum and globally defined phase.
 In the non-compact case, however, there generally is also a continuous part of the spectrum which is often probed by quantum scattering and transport. In these non-compact cases it appears straightforward to utilize the Bloch theorem and related concepts of crystal physics to describe the quantum behavior in what we shall call a \textit{corrugation crystal}. However,  this path has been rarely explored and only for special manifolds \citep{Aoki2001, Fujita2005} or in photonics \citep{Longhi2007, Szameit2010, Schultheiss2010}, and is often referred to as \textit{topological crystal} (although the effects are due to geometry rather than topology). The issue of topological effects  in this context has also been discussed \citep{Pandey2016, Pandey2018}. Here, we    rigorously formulate the  quantum mechanics of  a particle in corrugation crystals and present   explicit results and simulations for quasi-one-dimensional example. Already in this simple case we find nontrivial position-dependent diffusion (or mass) which is induced by geometry (rooted in the space  metric tensor), contributes to the band structure, and does not occur in usual crystals defined by a periodic scalar potential only. Linear transport is considered by calculating the transmittance following  the Landauer-Büttiker approach. \\
{\it{Model.-}}
We theoretically describe the quantum states and transport coefficients  of a particle constrained to a two-dimensional Riemannian manifold $(\mathcal{M},g)$ that is isometrically embedded in the three-dimensional ambient Euclidean space $(\mathbb{R}^{3},\delta)$. 
This means, $\mathrm{dim}(\mathcal{M}) = 2$. For clarity and simplicity, the curvature is present only along one of two orthogonal directions.
\begin{figure*}[!]
\centering
\includegraphics[width=0.8\textwidth]{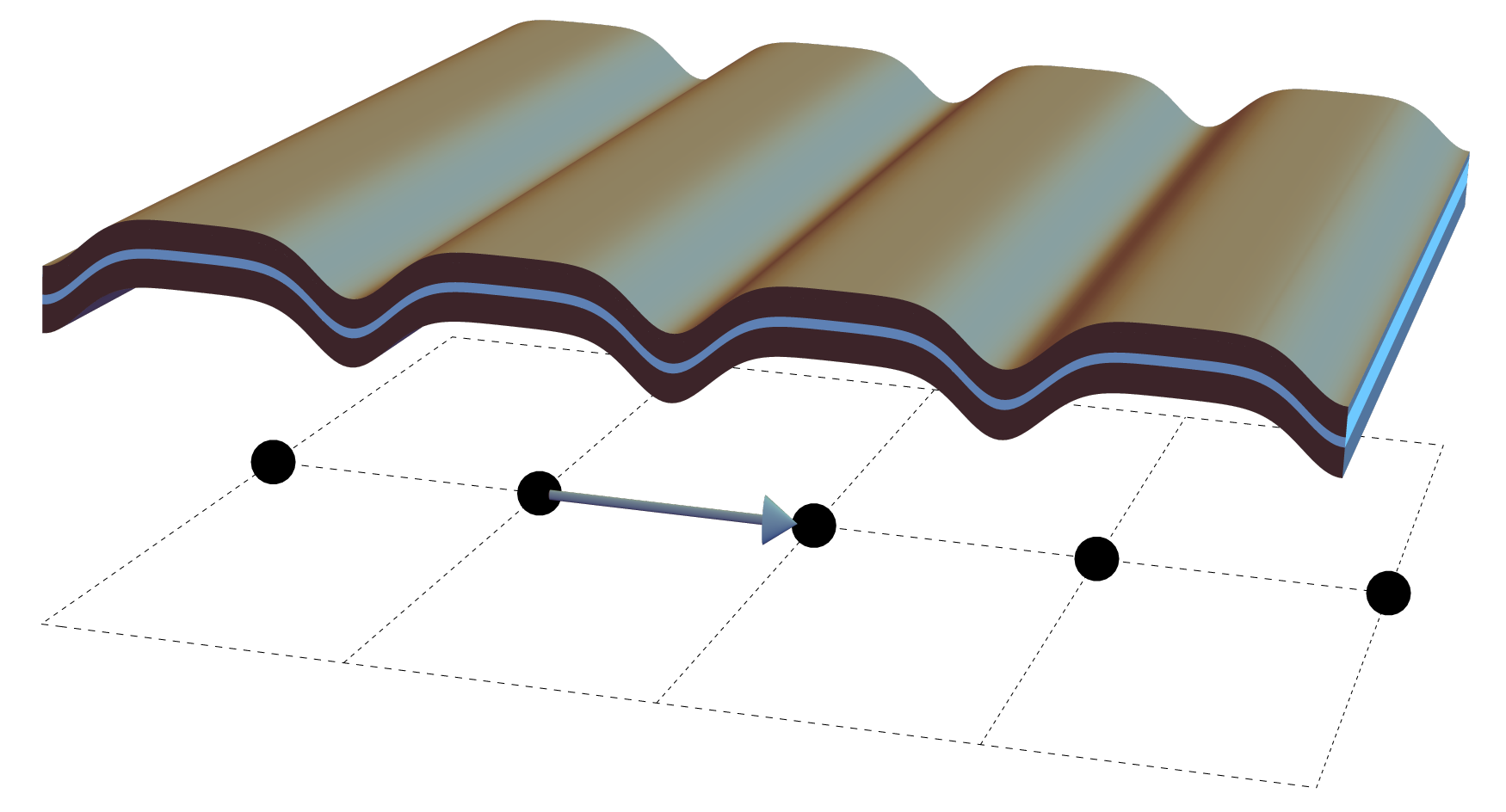}
\caption{Schematics of a two-dimensional Riemannian manifold $(\mathcal{M}, g)$ (marked blue), embedded in higher dimensional space (brown area) and  corrugated to a crystal in one dimension with a period $\Lambda^{1}$. The particle moves freely along $(\mathcal{M}, g)$.}
\label{fig:Scheme}
\end{figure*}
The metric tensor field $g$ is canonically defined as the inherited Euclidean inner product and the unique Levi-Civit\`{a} connection is assumed to act on the tangent bundle $T\mathcal{M}$. We study a space given as the trace of the Monge parametrization
\begin{align}
    \mathcal{X}_{\mathcal{M}}: \mathbb{R}^{2} \supset \mathcal{Q}_{\mathrm{t}} \to \mathcal{M} \subset \mathbb{R}^{3}\ ,\ \textbf{q} \mapsto \mathcal{X}_{\mathcal{M}}(\textbf{q}) \coloneqq \begin{pmatrix}
        q^{1} \\
        q^{2} \\
        f(q^{1})
    \end{pmatrix} = \textbf{x}\ ,
\label{eq:parametrization_Monge}
\end{align}
with $f\in C^{3}(\mathbb{R})$ (a schematics is shown by
Fig.\,\ref{fig:Scheme}). We target the case where the metric tensor is periodic which bears the expectation of the formation of a band structure $E_{\nu}(\mathbf{k})$, solely due to quantum and metric effects (note that a classical particle strictly constrained to the manifold moves along the geodesics and experiences no binding potential due to curvature). For well-defined $\mathbf{k}$ we can then follow the procedure of quantum and anomalous transport, as sketched above. We note, $\mathbf{k}$ here is a parameter representing a good (improper) quantum number and is not a dynamical variable. This is insofar important as the kinetic energy in our case is not diagonal in a plane-wave basis. It is also important to emphasize that we will be dealing with quantum ballistic transport in the sense of the Landauer-B\"uttiker formalism. Treatment of diffusive (Boltzmann-type) transport requires the knowledge of the classical-quantum correspondence of the equations of motion on general manifolds, an issue which still needs clarification.\\
{  Another  observation is that global topology of the manifold (not of the corrugation), for example if 
	$({\mathcal{M}}, g)$ is wrapped to a cylinder with a large radius and local corrugation, is manifested in global  response  functions, such as the Aharonov-Bohm current around the cylinder when it is pierced by a magnetic flux  along its axis. The magnitude of the current depends however on the local structure of $({\mathcal{M}}, g)$, since the effective vector potential is set by $g_{\mu\nu}$ due to gauge invariance \cite{Bendin2025}. Henceforth, for numerical simulations we specialize to non-compact manifold, however.}
\par
The  spatial periodicity  is embodied in $f(q^{1}) = f(q^{1} + \Lambda^{1})$ for all $q^{1}\in \mathbb{R}$ w.\,r.\,t. some $\Lambda^{1}\in \mathbb{R}^{+}$. From the parametrization, the canonical tangent vector fields are derived as
\begin{align}
    \mathrm{D}\mathcal{X_{\mathcal{M}}}(\textbf{q})
        =
        \begin{pmatrix}
            \textbf{t}_{1}(\textbf{q}) & \textbf{t}_{2}(\textbf{q})
        \end{pmatrix}
        =
        \begin{pmatrix}
            1 & 0 \\
            0 & 1 \\
            f_{,1}(q^{1}) & 0
        \end{pmatrix}\ ,
\end{align}
which fully characterize the metric
\begin{align}
    \left( g_{\mu\nu} \right)(\textbf{q}) = \begin{pmatrix}
        1 + f_{1}(q^{1})^{2} & 0 \\
        0 & 1
    \end{pmatrix}\ .
\label{eq:metric_tensor}
\end{align}
Constructing the orthogonal normal vector field completes the local frame such that a sufficiently small neighborhood $(\mathcal{N}_{\mathcal{M}},G)$ is parametrized by
\begin{align}
    \mathcal{X}: \mathbb{R}^{3} \supset \mathcal{Q}_{\mathrm{t}} \times\mathcal{Q}_{\mathrm{n}} \to \mathcal{N}_{\mathcal{M}} \subset \mathbb{R}^{3}\ ,\ (\textbf{q},y^{3}) \mapsto \mathcal{X}(\textbf{q},y^{3}) \coloneqq \mathcal{X}_{\mathcal{M}}(\textbf{q}) + y^{3}\,\textbf{n}_{3}(\textbf{q})\ .
\end{align}
\par
Performing the procedure of dimensional reduction known as \textit{confining potential approach} (CPA) \citep{Jensen1971, Costa1981, Maraner2008, Brandt2017, Meschede2023, Bendin2025} yields the equation of motion for the particle on $(\mathcal{M},g)$. In brief, the constraint is modeled by a confining potential $V_{\mathcal{M}}$ with the following properties: (i) $V_{\mathcal{M}}$ depends on the normal displacement coordinate only and is homogeneous along $\mathcal{M}$; (ii) $V_{\mathcal{M}}$ possesses a deep minimum on $\mathcal{M}$; (iii) $V_{\mathcal{M}}$ respects a sub-group of the isometry group of the ambient space which defines the gauge group of the resulting effective theory. This means, with weaker $V_{\mathcal{M}}$ we smoothly transition for a motion on $(\mathcal{M},g)$ to one in the hyperspace. Furthermore, the procedure assures that the Heisenberg uncertainty principle applies. In fact, the zero-point energy/motion set by $V_{\mathcal{M}}$ is key to differences between classical mechanical systems with constraints and quantum systems on manifolds. A usual good choice is to assume the confining potential as harmonic.
\par
Pulling the Schrödinger equation back into the parameter space on behalf of $\mathcal{X}$,
\begin{align}
    \ii \hbar\,\partial_{t}(\mathcal{X}^{\ast}\psi)(q,\textbf{y},t) = \left[ -\frac{\hbar^{2}}{2m}\,\nabla^{M}\nabla_{M} + V_{\mathcal{M}}(\textbf{y})\,\hat{I} \right](\mathcal{X}^{\ast}\psi)(q,\textbf{y},t)\ ,
\end{align}
and applying the CPA \citep{Costa1981, Meschede2023, Bendin2025}
yield the effective Schrödinger equation for a particle moving on the manifold $(\mathcal{M},g)$,
\begin{align}
    \ii\hbar\,\partial_{t}\chi_{\mathrm{t}}(\textbf{q},t) = -\frac{\hbar^{2}}{2m}\,\left[ \nabla^{\mu}\nabla_{\mu} + (M^{2}(\textbf{q}) - K(\textbf{q}))\,\hat{I} \right]\chi_{\mathrm{t}}(\textbf{q},t)\ .
    \label{eq:CPA_extrinsic_SE}
\end{align}
Here, the Laplace-Beltrami operator is given by the Voss-Weyl formula
\begin{align}
    \nabla^{\mu}\nabla_{\mu} = \frac{1}{\sqrt{\vert g\vert}}\,\partial_{\mu}\left( \sqrt{\vert g\vert}\,g^{\mu\nu}\,\partial_{\nu} \right)\ .
\label{eq:Laplace-Beltrami}
\end{align}
A geometry-induced contribution appears which is commonly called the geometric potential \citep{Wang2017},
\begin{align}
    V_{\mathrm{geo}} \coloneqq -\frac{\hbar^{2}}{2m}\,(M^{2} - K)\ .
\label{eq:geometric_potential}
\end{align}
$V_{\mathrm{geo}}$ depends on the mean ($M = \frac{1}{2}(\kappa_{1} + \kappa_{2})$) and Gaussian ($K = \kappa_{1}\kappa_{2}$) curvatures of the surface $\mathcal{M}$ (in terms of the principal curvatures $\kappa_{1}$ and $\kappa_{2}$). For nontrivial curvature along one direction only, e.\,g. $\kappa_{2} = 0$, one finds from \eqref{eq:parametrization_Monge}
\begin{align}
    V_{\mathrm{geo}}(q^{1}) = -\frac{\hbar^{2}}{8m}\,\kappa_{1}^{2}(q^{1}) = -\frac{\hbar^{2}}{8m}\,\frac{f_{,11}^{2}(q^{1})}{[ 1 + f_{,1}^{2}(q^{1}) ]^{3}}\ .
\end{align}
With this, the expression
\eqref{eq:CPA_extrinsic_SE} can be simplified using the separation ansatz $\chi_{\mathrm{t}}(\textbf{q},t) = \chi_{\mathrm{t}}^{(1)}(q^{1})\,\chi_{\mathrm{t}}^{(2)}(q^{2})\,\ee^{-\ii\frac{E}{\hbar}t}$ with $E = E^{(1)} + E^{(2)}$ resulting in
\begin{align}
\begin{aligned}
    -\frac{\hbar^{2}}{2m}\,\left[ \frac{1}{1 + f_{,1}^{2}}\partial_{1}^{2} - \frac{f_{,1}\,f_{,11}}{(1+ f_{,1}^{2})^{2}}\partial_{1} + \frac{1}{4}\,\frac{f_{,11}^{2}}{(q + f_{,1}^{2})^{3}}\,\hat{I} \right]\,\chi_{\mathrm{t}}^{(1)}(q^{1}) &= E^{(1)}\,\chi_{\mathrm{t}}^{(1)}(q^{1})\ , \\
    -\frac{\hbar^{2}}{2m}\partial_{2}^{2}\,\chi_{\mathrm{t}}^{(2)}(q^{2}) &= E^{(2)}\,\chi_{\mathrm{t}}^{(2)}(q^{2})\ .
\end{aligned}
\label{eq:CPA_extrinsic_SE_separated}
\end{align}
Even though the particle is free to move on the manifold, and we have no external scalar potential, the resulting equation of motion is crossly different from the flat-space case for both a free particle or a particle subject to a periodic scalar potential (conventional Bloch particle). To highlight the difference/analogy more, on can introduce \citep{Schwager2025} a geometry-dependent effective mass and  a scalar potential function as
\begin{align}
    m_{\mathrm{eff}}(q) \coloneqq \left( 1 + f_{,1}^{2}(q^{1}) \right)\,m = \Vert \textbf{t}_{1}(q^{1})\Vert_{2}^{2}\ m\ , 
\label{eq:effective_mass}
\end{align}
and
\begin{align}
    V_{\mathrm{eff}}(q^{1}) \coloneqq \left( 2\,\left( 1+f_{,1}^{2}(q^{1}) \right)\,\frac{f_{,1}(q^{1})\,f_{,111}(q^{1})}{f_{,11}^{2}(q^{1})} + 3\,\left( 1 - f_{,1}^{2}(q^{1}) \right) \right)\,V_{\mathrm{geo}}(q^{1})\ .
\label{eq:effective_potential}
\end{align}
The solution of the first equation in \eqref{eq:CPA_extrinsic_SE_separated} can be composed as
\begin{align}
    \chi_{\mathrm{t}}^{(1)}(q^{1}) \doteq \sqrt[4]{1 + f_{,1}^{2}(q^{1})}\,\varphi(q^{1}) = \sqrt{\Vert \textbf{t}_{1}(q^{1})\Vert_{2}}\,\varphi(q^{1})\ ,
\end{align}
where $\varphi$ is given as the solution of
\begin{align}
   \left[ -\frac{\hbar^{2}}{2\,m_{\mathrm{eff}}(q^{1})}\partial_{1}^{2} + V_{\mathrm{eff}}(q^{1})\,\hat{I} \right]\,\varphi(q^{1}) =  E^{(1)}\,\varphi(q^{1})\ .
\label{eq:SE_effective}
\end{align}
Equation \eqref{eq:SE_effective} is formally a usual one-dimensional stationary Schrödinger equation. An external potential field can be included by an appropriate mapping on the manifold \citep{Bendin2025}. E.\,g., it acts as a scalar potential added to $V_{\mathrm{geo}}$ contributing to $V_{\mathrm{eff}}$ but the structure of the equation remains unchanged.
\par
Due to the periodicity of $f$ all geometric dependencies and derived quantities \eqref{eq:metric_tensor}, \eqref{eq:Laplace-Beltrami} - \eqref{eq:geometric_potential}, and \eqref{eq:effective_mass} - \eqref{eq:effective_potential} are periodic, too. Mathematically, the Bloch theorem for the solution of the differential equation applies in the known way suggesting the ansatz
\begin{align}
    \varphi_{k^{1}}(q^{1}) \doteq \ee^{\ii k_{1}q^{1}}\,u_{k_{1}}(q^{1})\ .
\end{align}
One can then construct the reciprocal lattice within the $k^{1}$-parameter space. From \eqref{eq:SE_effective} the $k^{1}$-dependent energy eigenvalues follow which we can order in bands corresponding to the $\nu$-th Brillouin zone ($\nu \in \mathbb{N}$ is the band index),
\begin{align}
    \left[-\frac{\hbar^{2}}{2\,m_{\mathrm{eff}}(q^{1})}\left( \partial_{1}^{2} + \ii\,k^{1}\partial_{1} - (k^{1})^{2}\,\hat{I} \right) + V_{\mathrm{eff}}(q^{1})\,\hat{I} \right]\,u_{\nu, k^{1}}(q^{1}) = E_{\nu}^{(1)}(k^{1})\,u_{\nu,k^{1}}(q^{1})\ .
\label{eq:Bloch_problem}
\end{align}
{\it Bandstructure.-} The solutions of \eqref{eq:Bloch_problem} determine the admissible stationary quantum states that form on the periodically corrugated manifold $(\mathcal{M},g)$. The energy dispersion relation $E = E_{\nu}(k^{1})$ yields a band structure inside  one-dimensional reciprocal space, wherein the sizes of the Brillouin zones are determined by the curvature periodicity $2\pi/\Lambda^{1}$.
We recall, the spectrum is solely due to the curvature and geometry structure of the space.
As the particle is confined to move on a curved manifold, a spatially varying coefficient is introduced in the Laplace term. The latter can be understood as a second rank tensor potential, i.\,e., a  position-dependent effective mass acts like an inhomogeneous (onedimensional) diffusion tensor.
\par
Since quantum states are associated with our band structure in the parameter $\textbf{k}$-space, one can construct within some energy bandwidth  a wave packet and its   group velocity (for traversing the manifold) reads 
\begin{align}
    {\textbf{v}_{\mathrm{g}}}_{\nu}(\textbf{k}) = \frac{1}{\hbar}\,\boldsymbol{\nabla}_{\textbf{k}}E_{\nu}(\textbf{k})\ .
    \label{eq:group_velocity}
\end{align}
We can also introduce a spectral mass tensor as
\begin{align}
    M^{-1}_{\nu}(\textbf{k}) = \frac{1}{\hbar^{2}}\,\boldsymbol{\nabla}_{\textbf{k}} \otimes \boldsymbol{\nabla}_{\textbf{k}} E_{\nu}(\textbf{k})\ ,
\end{align}
which in our case, where {  $E(\textbf{k}) = E^{(1)}_{\nu}(k^{1}) + {(\hbar k^2)^{2}}/{(2m)}$,}
implies
{ 
\begin{align}
    (M^{ij}_{\nu})(k^{1}) = \begin{pmatrix}
        \hbar \left[\partial_{k^{1}} {v_{\mathrm{g}}}_{\nu\,1}\right]^{-1} & 0 \\
        0 & m
    \end{pmatrix}\ .
    \label{eq:spectral_mass}
\end{align}
}
%
The density of states incorporating the Kramers degeneracy of a particle with spin $s = {1}/{2}$ is given by 
\begin{align}
    D(E) = \frac{2}{(2\pi)^{2}}\,\sum\limits_{\nu\in\mathbb{N}}\,\int\limits_{\mathrm{1BZ}} \delta(E - E_{\nu}(\textbf{k}))\ \mathrm{d}^{2}\mathbf{k}\ ,
\end{align}
which can be reformulated as an integral over isoenergy surfaces,
\begin{align}
    D(E) = \frac{2}{(2\pi)^{2}}\,\sum\limits_{\nu\in\mathbb{N}}\,\int\limits_{E_{\nu}(\textbf{k}) = E} \frac{1}{\Vert \boldsymbol{\nabla}_{\mathbf{k}}E(\mathbf{k}) \Vert_{2}}\ \mathrm{d}S_{E}(\textbf{k})\ .
    \label{eq:DOS}
\end{align}
In our evaluation below, we will however consider it w.\,r.\,t. one dimension only.
\par
\begin{figure*}[!]
\centering
\includegraphics[width=0.8\textwidth]{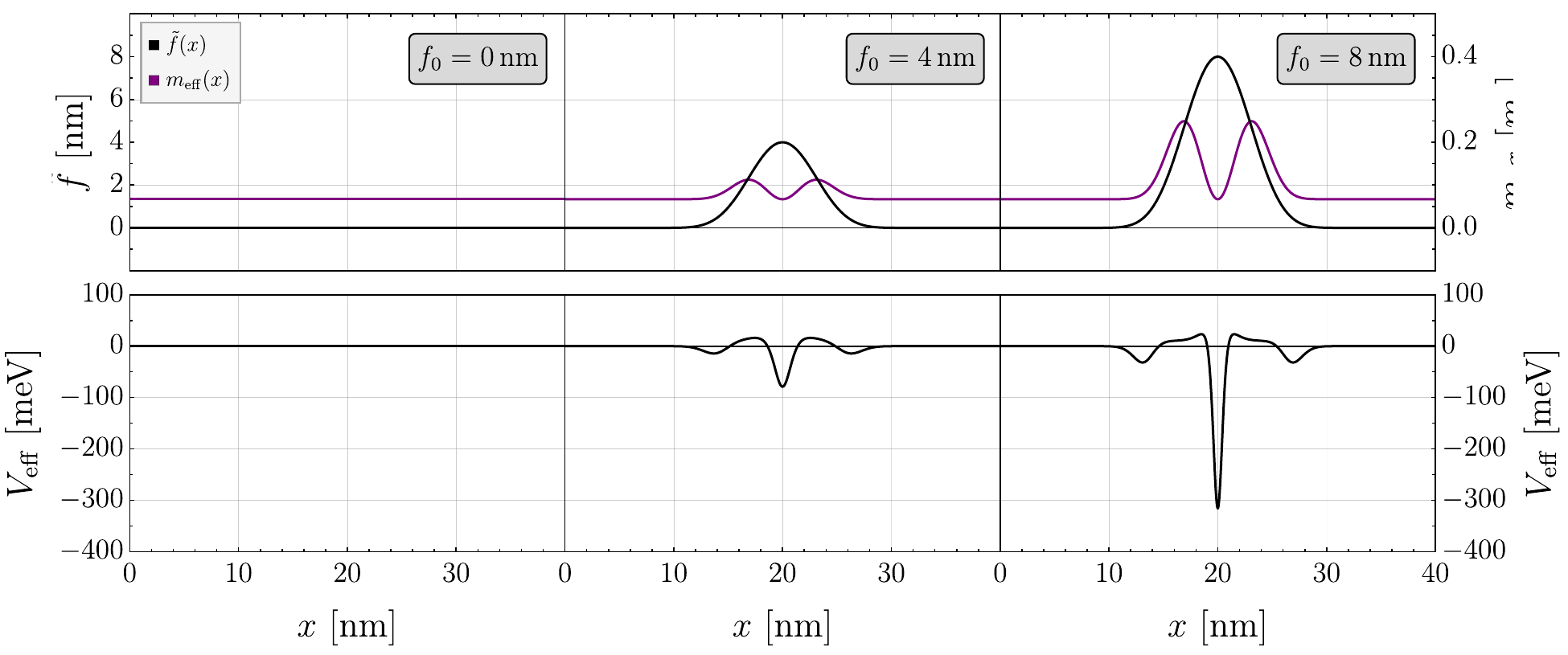}
\caption{Spatial course of the elementary dent described by \eqref{eq:dent_parametrix} as well as the geometry-induced effective mass and potential. The setup is characterized by the parameters  $m = 0.067\,m_{e}$, $\Lambda^{1}=40\ \mathrm{nm}$, $w=21.5$, and $q_{0}^{1}=20\ \mathrm{nm}$, while  $f_{0}$ increases across the rows. The leftmost row refers to flat space which is the reference system.}
\label{fig:Setting}
\end{figure*}

{\it Transport.-} Steady-state conductance through an asymptotically flat, periodically corrugated space can be described via the stationary S-matrix approach, which translates the incoming signal to the outgoing one. In one dimension, we assume that $N$ corrugations are placed between $q^{1}_{\mathrm{L}} = 0$ and $q^{1}_{\mathrm{R}} = N\,\Lambda^{1}$, serving as the scattering region. Considering a single corrugation ($N=1$) first, the wave function in the asymptotic region reads $\chi_{\mathrm{t}}^{(1)}(q^{1}) = A_{\mathrm{L}}\,\ee^{\ii k_{1}q^{1}} + B_{\mathrm{L}}\,\ee^{-\ii k_{1}q^{1}}$ for $q^{1} \leq 0$ and $\chi_{\mathrm{t}}^{(1)}(q^{1}) = A_{\mathrm{R}}\,\ee^{-\ii k_{1}q^{1}} + B_{\mathrm{R}}\,\ee^{\ii k_{1}q^{1}}$ for $q^{1}\geq \Lambda^{1}$, respectively. Then, we can map these basis states onto a matrix model and obtain
\begin{align}
    \begin{pmatrix}
        B_{\mathrm{L}} \\
        B_{\mathrm{R}}
    \end{pmatrix} = \begin{pmatrix}
        S_{11} & S_{12} \\
        S_{21} & S_{22}
    \end{pmatrix} \cdot \begin{pmatrix}
        A_{\mathrm{L}} \\
        A_{\mathrm{R}}
    \end{pmatrix}\ .
\end{align}
As all coefficients in \eqref{eq:CPA_extrinsic_SE_separated} are real the S-matrix is time-reversal symmetric, so due to its unitarity it is symmetric, $S^{T} = S$. Then, the reflectance and transmittance are identical for incoming signals from both sides and are given by
\begin{align}
    R = \vert S_{11}\vert^{2} = \vert S_{22}\vert^{2}\ ,\quad T = \vert S_{21}\vert^{1} = \vert S_{12}\vert^{2}\ ,
\end{align}
fulfilling $R + T = 1$. In addition, the transfer matrix translates the left side to the right side and is defined according to
\begin{align}
    \begin{pmatrix}
        B_{\mathrm{R}} \\
        A_{\mathrm{R}}
    \end{pmatrix}  =  \begin{pmatrix}
        M_{11} & M_{12} \\
        M_{21} & M_{22}
    \end{pmatrix} \cdot \begin{pmatrix}
        A_{\mathrm{L}} \\
        B_{\mathrm{L}}
    \end{pmatrix}\ .
\end{align}
It is possible to translate the S-matrix into the M-matrix and vice versa \citep{Reider2022}. Here, we exploit the transfer matrix approach to compute the total scattering of multiple adjacent corrugations as the total M-matrix is determined by (left-)multiplication,
\begin{align}
    M^{\mathrm{tot}} = \prod\limits_{i=1}^{N} M^{(i)}\ .
\end{align}
%
%
{\it Numerical results:-}
For numerical implementation, the deformation is assumed to be a periodic repetition of the bump function $\tilde{f} \in C_{c}^{\infty}\left( \Omega_{1\mathrm{WSZ}} \right)$, with the first Wigner-Seitz cell $\Omega_{1\mathrm{WSZ}} = (0, \Lambda^{1})$,
{ 
\begin{align}
    \tilde{f}(q^{1}) \coloneqq f_{0 } \begin{cases} 
    \mathrm{exp}\left\{ w\,\left[ 1 + 
    \left[ \frac{q^{1} - q^{1}_{0}}{\Lambda^{1} - q^{1}_{0}} \right)^{2} - 1 \right]^{-1} \right\}
 &: \vert q^{1} - q^{1}_{0} \vert < \Lambda^{1} - q^{1}_{0} ,\\
    0 &: \vert q^{1} - q^{1}_{0} \vert \geq \Lambda^{1} - q^{1}_{0}.
    \end{cases}\
\label{eq:dent_parametrix}
\end{align}
}The parameters $f_{0},w\in \mathbb{R}^{+}$ describe the peak amplitude and the width of the bump, respectively, and $q^{1}_{0} \in \Omega_{1\mathrm{WSZ}}$ is its center. 
{   As relevant settings for an experimental realization we mention  the experiments on nanoscale geometry control of clamped graphene  membrane \cite{Alba2016} as well as   on WSe$_2$ locally dented from beneath  by SiO$_2$ nanopillars  \cite{Parto2021}
	or a strained monolayer of WS$_2$ locally nanoscale dented by an AFM tip \citep{Harats2020}.  Further methods on 
strained two-dimensional materials \citep{Peng2020,10.10631.3636410} can be useful.  As for the energy scale, the
binding energy of nanoscale curvature-induced states is found to be 
on the $\mathrm{meV}$-scale  \cite{Bendin2025}, measured from the Fermi level. Hence, the geometry-induced bound states discussed  here are  quite extended and  might be relevant for reactions on surfaces, optical properties, and transport. 
Space modulations may also be achieved  via engineered  gating of a semiconductor-based two-dimensional electron gas. Here, we are interested in a general demonstration of the geometry-induced phenomena and   adopt the parameters $m = 0.067\,m_{e}$, $\Lambda^{1} = 40\,\mathrm{nm}$, $w = 21.5$, and $q^{1}_{0} = 20\,\mathrm{nm}$, which are appropriate for $\text{Al}_{x}\text{Ga}_{1-x}\text{As}/\text{GaAs}$, and choose $f_{0} \in \{0; 4; 8\}\,\mathrm{nm}$. Typical  course of the corrugations, the effective mass \eqref{eq:effective_mass} and the effective potential $\eqref{eq:effective_potential}$ are shown in fig.\,\ref{fig:Setting}. }\\
\begin{figure*}[!]
\centering
\includegraphics[width=0.8\textwidth]{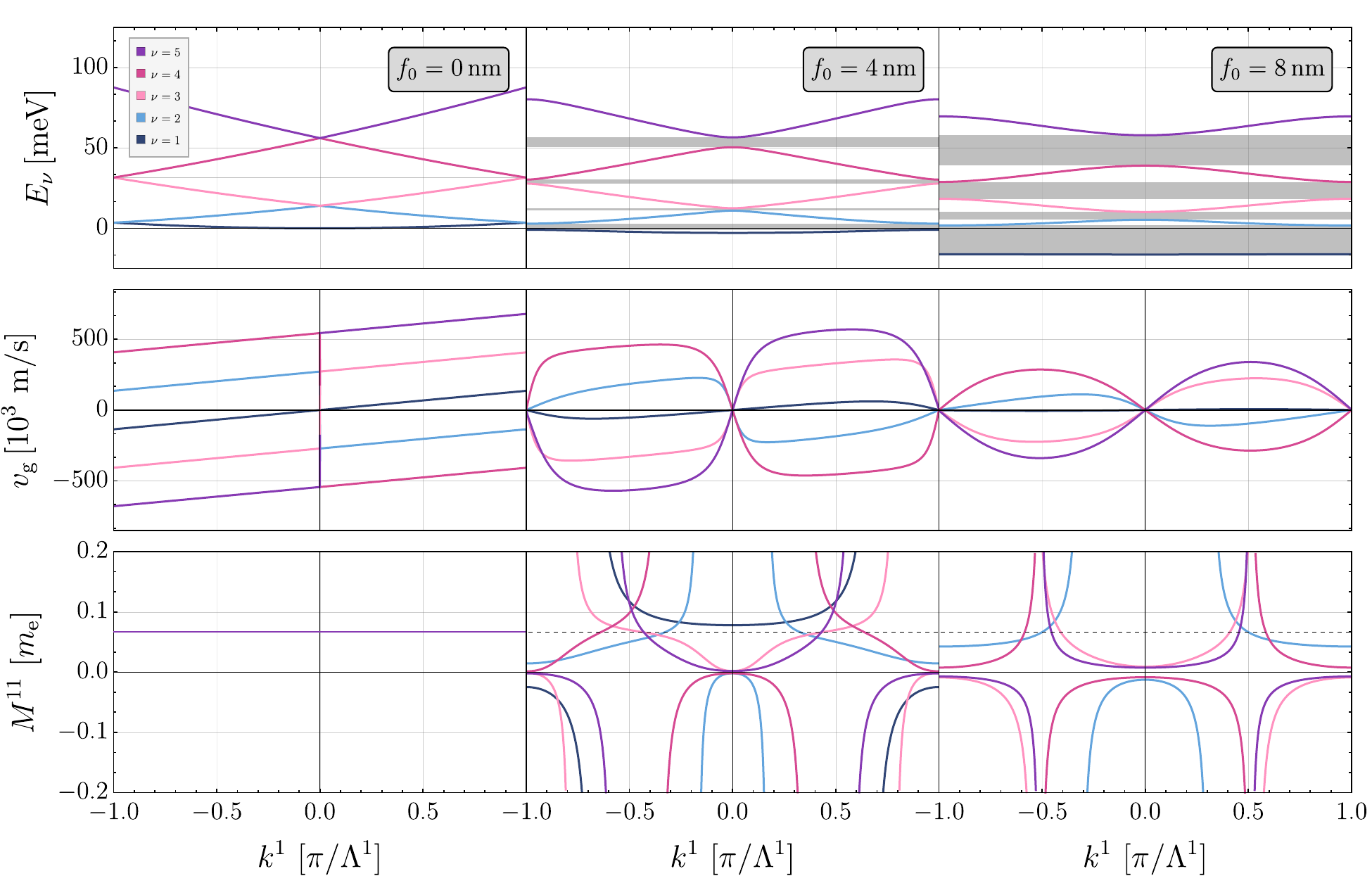}
\caption{Band structure, group velocity and spectral mass as function of $k^1$ for the 
parameters given  in fig.\,\ref{fig:Setting}, with $f_{0}$ increasing across the rows. 
Different bands are color coded. Band gaps are marked with gray shaded areas.
Leftmost row refers to  free electron gas in a flat space as reference system.}
\label{fig:Crystal_Data}
\end{figure*}
\begin{figure*}[!]
\centering
\includegraphics[width=0.8\textwidth]{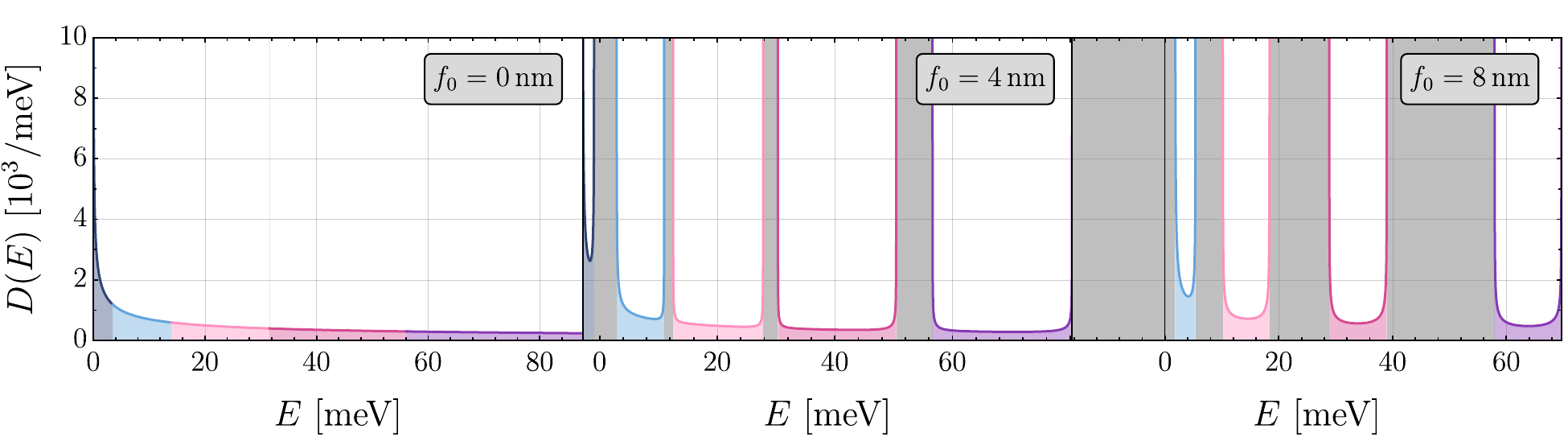}
\caption{Density of states for the corrugation crystals with the same parameters as in fig.\,\ref{fig:Setting}, with $f_{0}$ increasing across the rows. Only one direction is incorporated. Different bands are indicated by color. Spectral regions with gray shade do no host any states and correspond to the band gaps. Leftmost row is for  free electron gas.}
\label{fig:DOS}
\end{figure*}
{\it Crystal Structure.-} Fig.\,\ref{fig:Crystal_Data} compares the band structure, the group velocity \eqref{eq:group_velocity} and the spectral mass parameter \eqref{eq:spectral_mass} of the quasi-onedimensional corrugation crystal for the three different amplitudes. Fig\,\ref{fig:DOS} additionally shows the respective density of states \eqref{eq:DOS}. The band structure and the derived quantities are symmetric with respect to the center of the first Brillouin zone (i\,e. the $\Gamma$ point). In our model, the free electron gas in a flat space serves as the reference system. It is achieved when $f_{0} = 0$, so the leftmost columns in these plots show the corresponding familiar results. Keeping $w = \mathrm{const}.$ and increasing the amplitude $f_{0} > 0$ produces corrugations which correspond to nontrivial and increasing geometry-induced effective masses and potentials. Hence, the system increasingly deviates from the reference system as visible in all the observed quantities.
\par
The corrugations leads to a decrease of the energy values accompanied by a decrease of the band widths and the opening of band gaps, both lying in the order of $\approx \mathcal{O}(10\,\mathrm{meV})$. Thereby, the decrease of the band widths leads to the formation of flat bands for larger amplitudes $f_{0}$. Furthermore, for any $f_{0} > 0$ the lowest energy band is partially negative (until it is entirely negative). We interpret this as a band of bound states. Recalling   that for  one-dimensional system all corrugations 
lead to  formation of attractive potentials that possess at least one bound state \citep{Exner1989, Goldstone1992}, this bound energy band aligns with the picture of hybridizing bound  orbitals, with  orbitals being  mainly localized around the minimum of the effective geometric potential. The flattening of the band structure is accompanied by a decrease of the group velocities which here have a magnitude of $\approx \mathcal{O}(100\,\frac{\mathrm{km}}{s})$ and vanish at both the center and edges of the first Brillouin zone ($\Gamma$- and $X$-points, respectively).
\par
The spectral mass derived from the band structure means an effective parameter of the particle moving inside the corrugation crystal, along the $x$-direction. Its magnitude is set by the particle mass parameter $m$, but modifications arise due to the presence of corrugations. All bands but the first one yield the same value $M_{\nu}^{11}(k^{1}) = m$ at exactly two points within the first Brillouin zone. In general, the spectral mass is very small near the center ($\Gamma$) and the edges ($X$) of the Brillouin zone, and diverges at exactly two points in between where the group velocity is extremal. On both sides of these vertical asymptotes the sign of the spectral mass changes.
\par
In accordance with the opening of band gaps, the density of states is modified.  As evident in fig.\,\ref{fig:DOS}, the flatter the bands, the narrower is the spectral region where the states localize. 
Thus, the densities of states corresponding to the single bands are all increasing with $f_{0}$. 
Simultaneously, the spectral regions that do not host any states (i.\,e. the band gap regions) are widening. As expected for a quasi-1D crystal, the density of states diverges at the edges of the first Brillouin zone.\\
%
{\it Transport:-}
\begin{figure*}[!]
\centering
\includegraphics[width=0.8\textwidth]{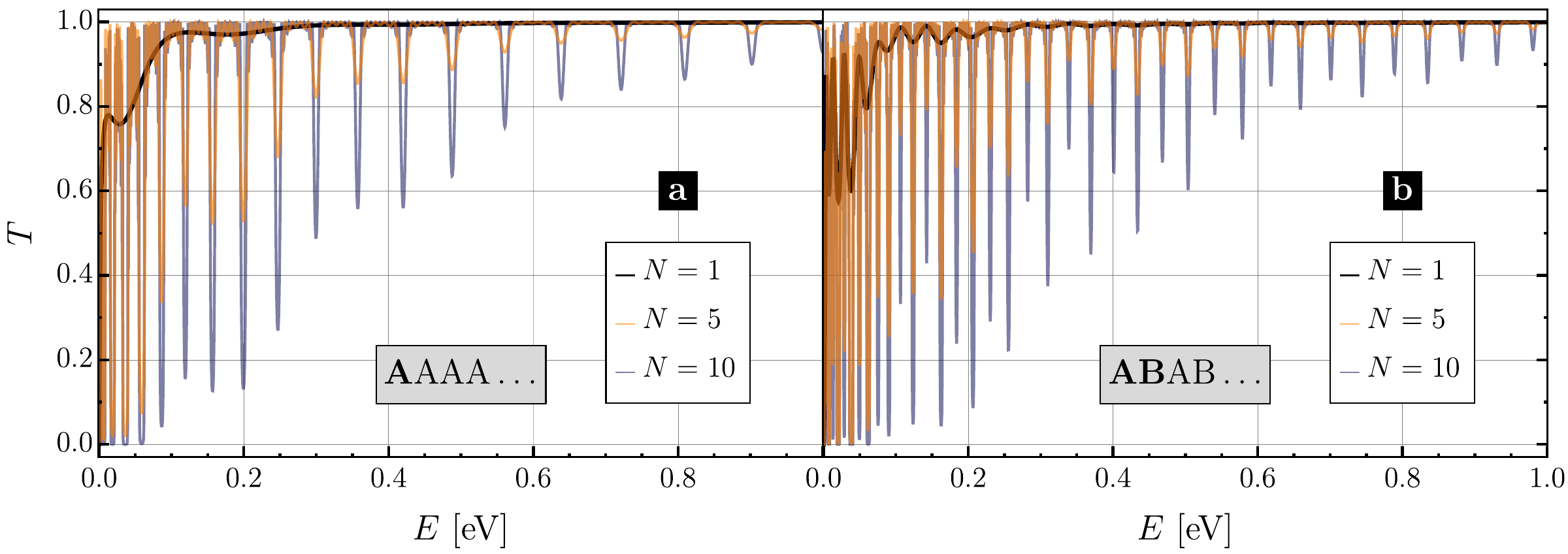}
\caption{Transmittance of a quantum particle traversing a $N$ fold corrugated space. The basis of the repeating structure is formed by corrugations with parameters $m = 0.067\,m_{e}$, $\Lambda^{1} = 40\,\mathrm{nm}$, $q_{0}^{1} = 20\,\mathrm{nm}$, $f_{0} = 8\,\mathrm{nm}$, as well as $w = 21.5$ (dent type A) or $w = 10.0$ (dent type B). (a) Transmittance for a simple lattice structure with basis A. (b) Transmittance for a sub-lattice structure with basis AB.}
\label{fig:Transport_Lattices}
\end{figure*}
To complement the band structure point of view we also simulated the transport properties of a quantum particle undergoing quantum scattering while traversing an approximate corrugation crystal of $N$ subsequent dents, using the transfer matrix approach.
For this purpose, we chose the parameters $m = 0.067\,m_{e}$, $\Lambda^{1} = 40\ \mathrm{nm}$, and $q_{0}^{1} = 20\ \mathrm{nm}$ as above, but fixing $f_{0} = 8\ \mathrm{nm}$. Instead, the width parameter $w$ is varied to illustrate the effects of sub-lattice structures and impurities.
\par
Fig.\,\ref{fig:Transport_Lattices}\,(a) shows the transmittance of a particle traversing $N$ identical corrugations with $w = 21.5$ (corresponding to the structure shown in the right-most rows of figs.\,\ref{fig:Setting}-\ref{fig:DOS}). The black line shows in the course of the transmittance $T$ for a single dent, which is almost fully converged to its limiting value one when the incidence energy surpasses $0.5\ \mathrm{eV}$. Note that the convergence $T \to 1\ ,\ (E\to +\infty)$ is an expected result for a infinitely extended scattering region due to the fundamental physics of quantum scattering processes \citep{Griffiths2004, Reider2022}. This curve agrees with the results found in similar studies previously reported in the literature \citep{Cao2019, Serafim2021} and by ourselves \citep{Schwager2025}.
\par
When the scattering region is extended by repeating corrugations, a series of resonances emerges. These are indicated by negative peaks which lower down up to $T = 0$ as $N \to +\infty$, and broaden out to some extent. Hence, they indicate spectral regions where perfect reflectance occurs. In between these, $T \approx 1$ is found, modulated by some ringing phenomenon. This is because  the transmittance $T$ obtains a rectangular shape when the corrugation infinitely many times repeats as $N\to +\infty$, jumping between $T=0$ and $T=1$. 
This behavior resembles the band structure of the corrugation crystal shown in figs.\,\ref{fig:Crystal_Data} and \ref{fig:DOS} as the band gaps mean spectral regions where no quantum states exist. 
This behavior   underlines the possible applicability of deformed wires as  band pass filters or reflectors for particle obeying the above equation of motion. In our case (assuming some GaAs heterostrucure), the data shows that for the lowest energy bands (compare fig.\,\ref{fig:Crystal_Data} and fig\,\ref{fig:Transport_Lattices}\,(a)) the limit of an infinitely extended crystal is reached in good approximation after ten repeating corrugations.
\par
To contrast the picture we now choose every second dent to possesses a width parameter of $w = 10.0$ rather than $w = 21.1$. In the language of crystal physics, these two corrugation form a sub-lattice structure according to the pattern $ABAB\ldots$, and a repetition of this two-component basis yields the transmittance data shown in fig.\,\ref{fig:Transport_Lattices}\,(b). The direct comparison with the one-component basis fig.\,\ref{fig:Transport_Lattices}\,(a) shows that the sub-lattice structure is evident  as a second series of negative peaks alternating with the previous ones. This behavior also agrees with previous studies \citep{Cao2019}.
\par
\begin{figure*}[!]
\centering
\includegraphics[width=0.5\textwidth]{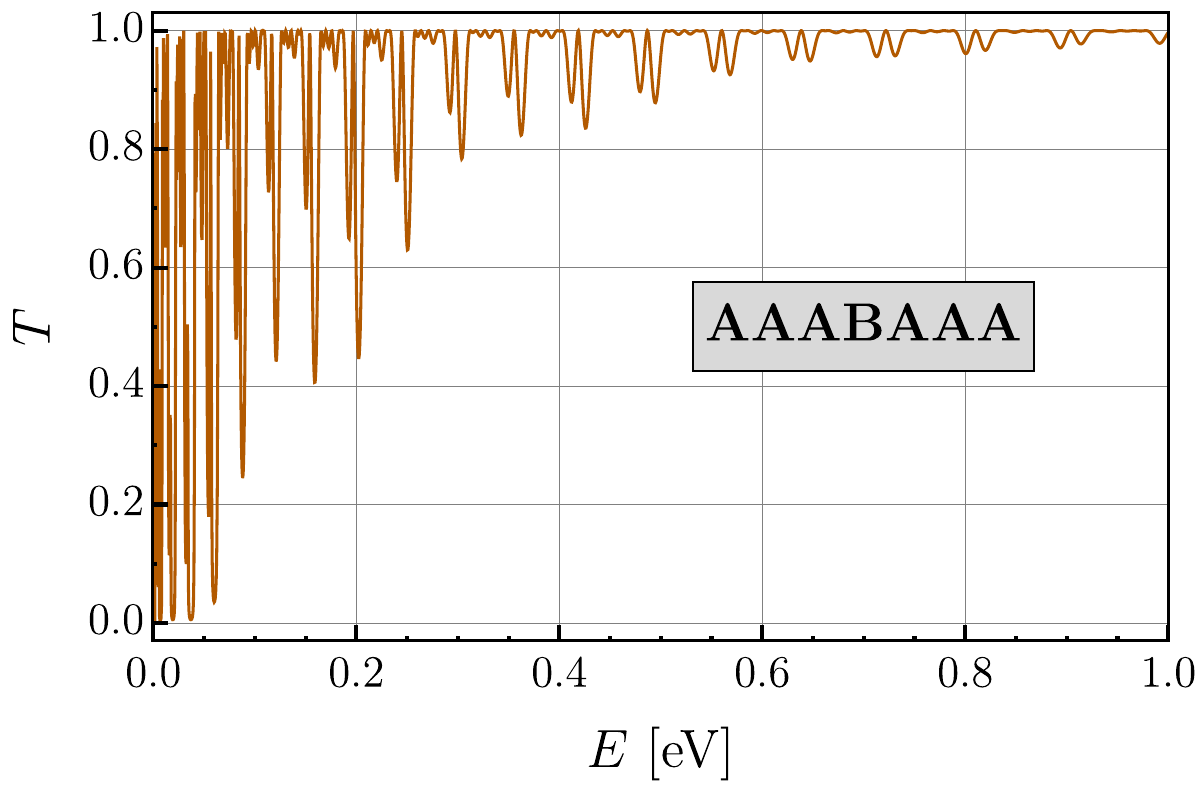}
\caption{Transmittance of a quantum particle traversing a corrugated space with impurity modeled by the dents described in fig.\,\ref{fig:Transport_Lattices}.}
\label{fig:Transport_Impurity}
\end{figure*}

Furthermore, comparing the black curves in figs.\,\ref{fig:Transport_Lattices}\,(a, b), which stand for one repetition of the respective corrugations only, shows that the transmittance is generally quite sensitive to the shape of the wire. This is empirically expected as the coefficients in the Schrödinger equation \eqref{eq:CPA_extrinsic_SE_separated} or \eqref{eq:SE_effective}, i.\,e. the geometry-induced the effective mass and potential, are nonlinearly dependent on the shape function $f$. To illustrate this even further, we utilized the second dent with $w=10.0$ as an impurity in a series of dents with the original shape for $w = 21.5$ and simulated the scattering process. The resulting transmittance for a structure of seven corrugations with the impurity in the center is shown in fig.\,\ref{fig:Transport_Impurity}. The result shows that the negative peaks split up into two. Thus, the effect of impurities is expected to be visible in the scattering data when the spectral resolution is high enough. Conversely, this allows the conclusion that  the potential of using conductance measurements for shape analysis of (periodically corrugated) quantum wires.
%
%

{\it Conclusions:-}
A constrained motion of a quantum particle  differs qualitatively   from its classical counterpart and from the motion of a quantum particle in flat space subject to a scalar potential. This is shown here for a periodically corrugated space and contrasted with 
a Bloch particle in flat space.  
In the case discussed here, the  influence of the metric tensor of the configuration space can be mapped onto 
Schrödinger-type equation with periodic geometry-induced potential  
and  an effective local mass parameter. We computed an exemplarily band structure and investigated the  group velocity, spectral mass, and density of states.
Based on the band structure, one can calculate the $\mathbf{k}$-space quantum geometric tensor and  Berry's curvature for studying quantum transport and for classifying states. This approach is (strictly) valid for effective single particle problems and is applicable here straightforwardly because of our special setting.
In at least two dimensions, even for a single particle, the metric tensor field $g$ is not necessarily diagonal. Then, the hyperspace axes and their associated reciprocal space would not yield independent quantum numbers as the $\textbf{k}$-variables would mix, prohibiting separation. This could lead to more sophisticated implications due to the nontrivial metric, but analyzing and classifying   this situation remains a task for future research.  On the other hand, once the solution  of the coupled equations on the manifold are found, one can span the projective Hilbert space and 
study geometric effects as discussed above. Regarding transport, 
for non-compact manifolds, the S-matrix theory we used here to study quantum transport still applies 
as long as the curved part of the space is limited and surrounded by   flat region  which  allows for the the definition of asymptotic scattering states that are needed to infer the conductance from the stationary scattering theory.
\par

\section*{Acknowledgment:}
This work has been supported by the Deutsche Forschungsgemeinschaft project number 429194455.
\section*{Competing Interests:}
The authors declare there are no competing interests.
\section*{Data Availability Statement:}
Necessary steps to arrive at the results of this work are provided in the text.

\bibliography{CPA_crystal}
\end{document}